\documentclass[a4paper,11pt]{article}
\usepackage{pos}
\usepackage{pdfpages}

\title{With complex Langevin towards the QCD phase diagram}
\ShortTitle{With complex Langevin towards the QCD phase diagram}

\author[a]{Felipe Attanasio}
\author*[b]{Benjamin Jäger}
\author[b,c]{Felix P.~G. Ziegler}

\affiliation[a]{Institute for Theoretical Physics, Universit\"{a}t Heidelberg, Philosophenweg 16, D-69120, Germany}

\affiliation[b]{CP3-Origins \& Danish IAS, Department of Mathematics and Computer Science, University of Southern Denmark, 5230 Odense, Denmark}

\affiliation[c]{
School of Physics and Astronomy, The University of Edinburgh, Edinburgh EH9 3FD, United Kingdom}

\emailAdd{pyfelipe@thphys.uni-heidelberg.de}
\emailAdd{jaeger@cp3.sdu.dk}
\emailAdd{felix.ziegler@ed.ac.uk}

\abstract{
We use complex Langevin simulations to explore the QCD phase diagram over a large range of chemical potentials and temperatures. For our simulations we use two flavours of dynamical Wilson fermions with a pion mass of approximately $480\,$MeV with a spatial volume of $24^3$. Here we report on consistency checks at zero chemical potentials and present our results for the fermion density and the Polyakov loop. We find that at the lowest temperature the fermion density remains zero until $m_N/3$, in line with the expectations from the Silver Blaze phenomenon. 
}

\FullConference{%
 The 38th International Symposium on Lattice Field Theory, LATTICE2021
  26th-30th July, 2021
  Zoom/Gather@Massachusetts Institute of Technology
}

\begin{document}
\maketitle
 
\section{Introduction}

The complex Langevin method (CL), first proposed by Parisi~\cite{Parisi:1980ys} has been successfully adapted and used in various models and approximations of QCD, for instance, the heavy-dense limit of QCD~\cite{Aarts:2016qrv}. Early results have been obtained for QCD with very heavy quarks~\cite{Sexty:2013ica,Scherzer:2020kiu}. Here, we aim to study the QCD phase diagram with significantly lighter quarks ($m_\pi \sim 480\,$MeV). Our simulations span a large range of chemical potentials, ranging up to approximately $6500\,$MeV, and various temperatures, as low as $25\,$MeV. 

\section{Complex Langevin}

The (complex) Langevin method evolves the gauge links of a lattice simulation along a new fictitious time $\theta$ for a small step size $\epsilon$. A first order update scheme can be written as 
\begin{align*}
    U_{x\mu} (\theta + \epsilon) = \exp \Big( \mathrm{i}\, \lambda^a\, ( - \epsilon \, D^a_{x\mu} S[U] + \sqrt{\epsilon} \, \eta_{x,\mu}^a)
    \Big) \, U_{x\mu} (\theta),
\end{align*}
where $\lambda^a$ are the Gell-Mann matrices, $D^a_{x\mu} S$ are the derivatives of the action with respect to the gauge link $U_{x\mu}$ and $\eta_{x,\mu}^a$ are the random white noise. The complexification is introduced by enlarging the group manifold from SU$(3)$ to SL$(3,\mathbb{C})$, i.e. by allowing the coefficients of the generators to become complex numbers
\begin{align*}
U_{x\mu} = \exp \Big( \mathrm{i}\, \lambda^a (A_{x\mu}^a + \mathrm{i}\,  B_{x\mu}^a) \Big). 
\end{align*}
The new time direction created by this procedure is analogue to the Molecular Dynamics time in standard lattice calculations. For sufficient long simulations, an observable $O$ can be obtained by taking an average over the Langevin time $\theta$
\begin{align*}
\langle O \rangle = \frac{1}{\theta_\mathrm{max} - \theta_\mathrm{therm}} \sum_{\theta = \theta_\mathrm{therm}}^{\theta_\mathrm{max}}  O[U(\theta)],
\end{align*}
after discarding a sufficient amount $\theta_\mathrm{therm}$ to remove thermalisation effects. As we aim to take the limit of $\epsilon \to 0$, proper treatment of autocorrelation effects is important. We use the automatic autocorrelation method presented in~\cite{Wolff:2003sm}. 

\section{Setup}

For our study we use two dynamical flavours of Wilson quarks without a clover term $c_{\mathrm{SW}} = 0$. Our simulation parameters are based on one of the setups used in~\cite{DelDebbio:2005qa}. The gauge coupling is fixed at $\beta = 5.8$ throughout all simulations. This implies a fixed lattice spacing of approximately $a\sim 0.06\,$fm~\cite{DelDebbio:2005qa}. The hopping parameter of $\kappa = 0.1544$ leads to a pion mass of $a\, m_\pi = 0.1458(7)$, which was measured using a HMC simulation at vanishing chemical potential using $N_t =128$. Converting the pion mass into physical units, we find $m_\pi \sim  480\,$MeV. The spatial simulation volume is fixed to $24^3$ and so the product $m_\pi L = 3.5$ is sufficiently large, so that volume effects are expected to be small. The authors of~\cite{DelDebbio:2005qa} found a pion mass of $a\, m_\pi = 0.1481(11)$ for a larger spatial volume of $32^3$. The difference is $1.5\%$, which indicates small volume effects. Different temperatures are realised by varying the temporal lattice extent. In this study, we use up to $128$ points in time, resulting in temperature down to $25\,$MeV. The chemical potential was scaled up to $a\, \mu = 2.0$. 
\begin{table}[h!]
    \centering
    \begin{tabular}{|c|c|}
         
         \hline
         $\beta = 5.8$  & $\kappa= 0.1544$ \\
         \hline
         $V = 24^3$ & $m_\pi L =3.5$\\
         \hline
         $N_t = 4 - 128$ & $T \sim 800 - 25\,$MeV  \\
         \hline
         $a\mu = 0.0 - 2.0 $ & $\mu \sim 0 - 6500\,$MeV  \\
                  \hline

    \end{tabular}
    \caption{The lattice parameters used in this study.}
    \label{tab:params}
\end{table}
Table~\ref{tab:params} shows a summary of the key parameters. To improve the convergence of the complex Langevin simulation, we employ adaptive step size scaling~\cite{Aarts:2009dg}, gauge cooling~\cite{Seiler:2012wz} and dynamic stabilisation~\cite{Attanasio:2018rtq,Attanasio:2020spv}.

\section{Extrapolation at zero chemical potential}

As a first consistency check, we compare HMC simulations with complex Langevin simulations at vanishing chemical potential. Even though this is a purely real setup and complex Langevin simulations are not necessary, we allow and start with configurations that are in the SL$(3,\mathbb{C})$ manifold. As complex Langevin suffers from finite step size corrections, we simulate multiple values of the step size $\epsilon$ and extrapolate to zero. As we employ step size scaling~\cite{Aarts:2009dg}, we measure the average step size $\langle\epsilon\rangle$ and adapt the relevant control parameter such that we achieve different values. \begin{figure}[!htb]
    \centering
    \begin{minipage}{.49\textwidth}
    \centering
    \includegraphics[width=0.98\textwidth]{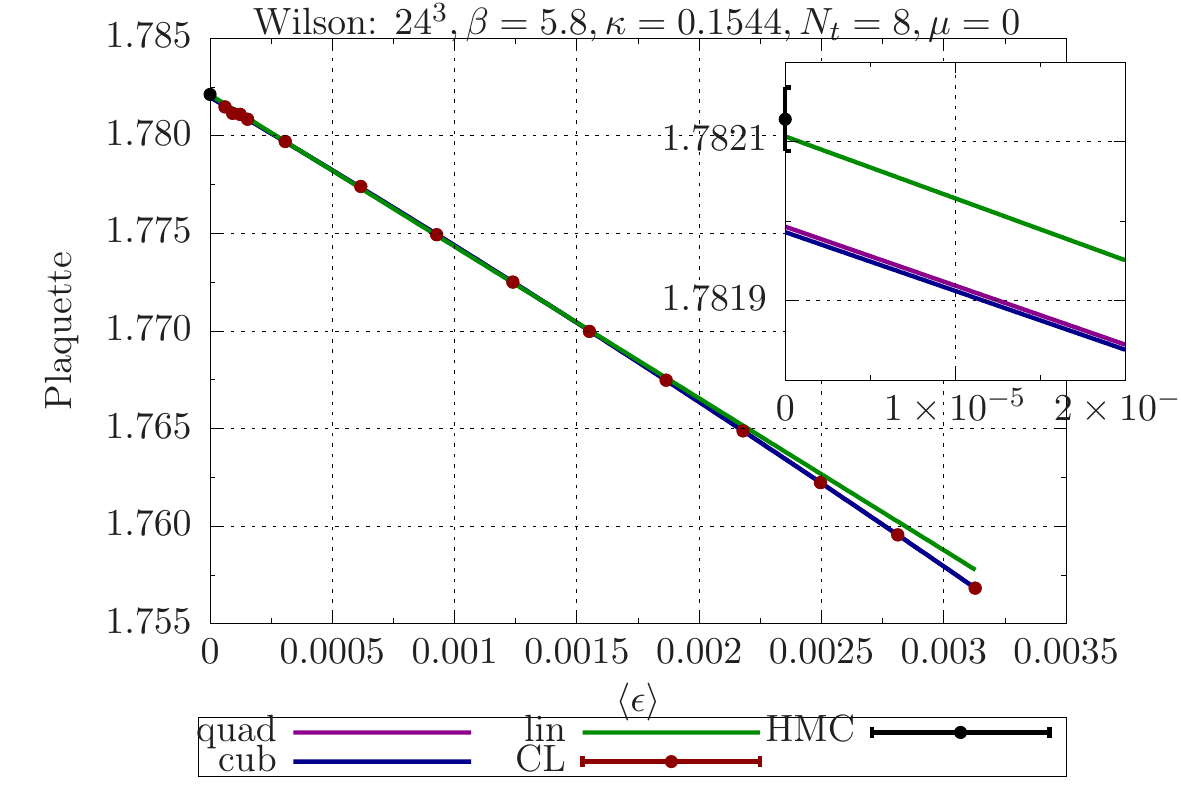}
    \end{minipage}
        \begin{minipage}{.49\textwidth}
    \centering
    \includegraphics[width=0.98\textwidth]{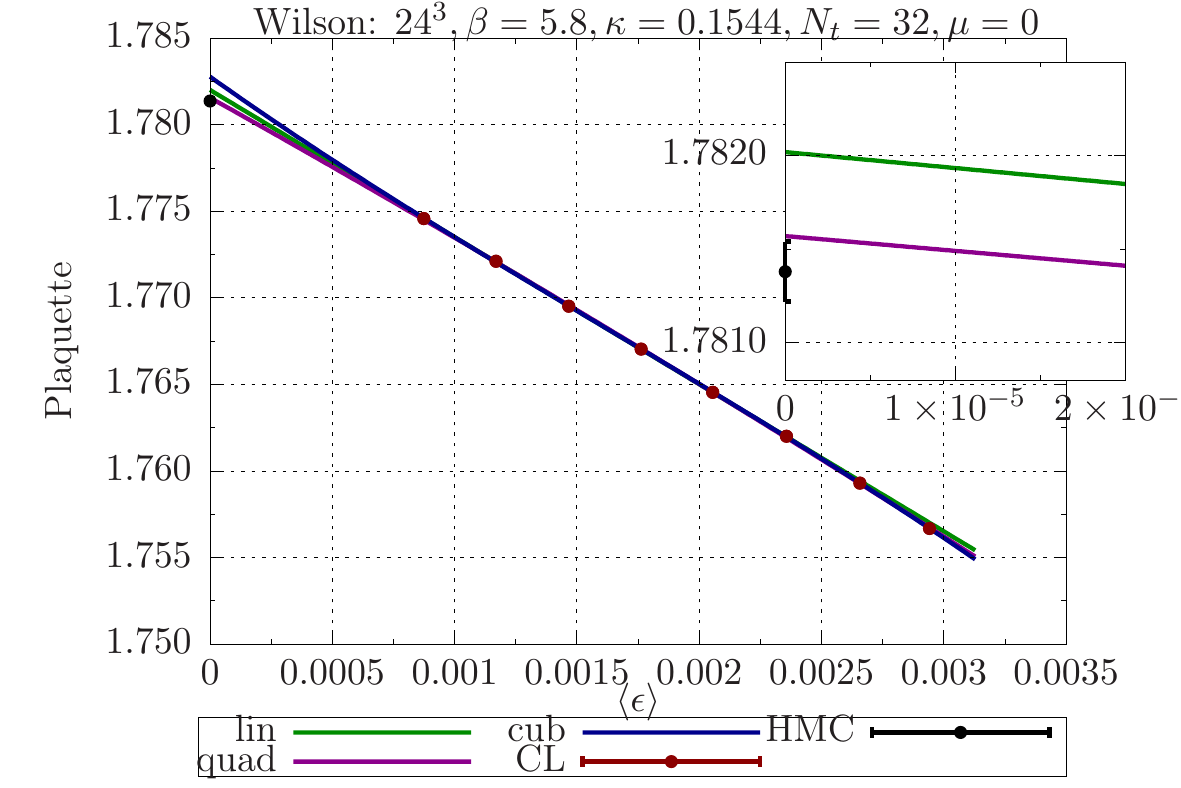}
    \end{minipage}
        \caption{The plaquette as function of the average Langevin step size $\langle\epsilon\rangle$ at zero chemical potential. The lines are different extrapolations towards the HMC result. (Left): High temperature regime, i.e. $N_t = 8$ corresponding to $T \sim 400\,$MeV. (Right): Low temperature regime, i.e. $N_t = 32$ corresponding to $T \sim 100\,$MeV}
    \label{fig:extra}
    
\end{figure}
Figure~\ref{fig:extra} shows the results for the plaquette as a function of the average step sizes for two different setups. The first setup (left panel) corresponds to large temperatures of approximately $400\,$MeV. The second setup (right panel) is at lower temperatures of approximately $100\,$MeV. The data points are augmented by three different fits to the data: a linear (lin), quadratic (quad) and cubic (cub) extrapolation. In the linear extrapolation, the four right most points have been removed for the fitting procedure. As we use a first-order integration scheme, we expect our step size correction to be of first order. This behaviour is clearly visible. Overall, we find very good agreement between the HMC results and the extrapolated result from complex Langevin simulations. The agreement is up to the 4th significant figure or less than a permille. 

\section{Simulation at non-zero chemical potential}

\begin{figure}[!htb]
    \centering
    \begin{minipage}{.49\textwidth}
    \centering
    \includegraphics[width=0.98\textwidth]{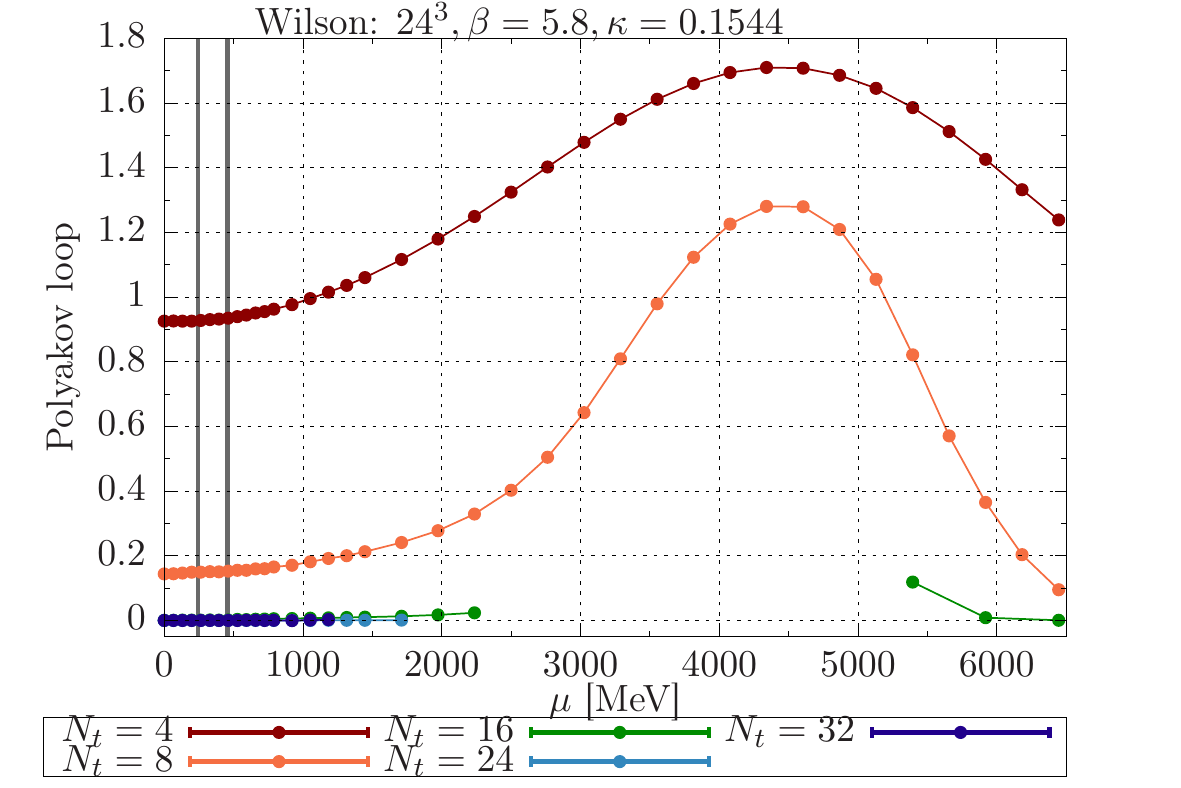}
    \end{minipage}
        \begin{minipage}{.49\textwidth}
    \centering
    \includegraphics[width=0.98\textwidth]{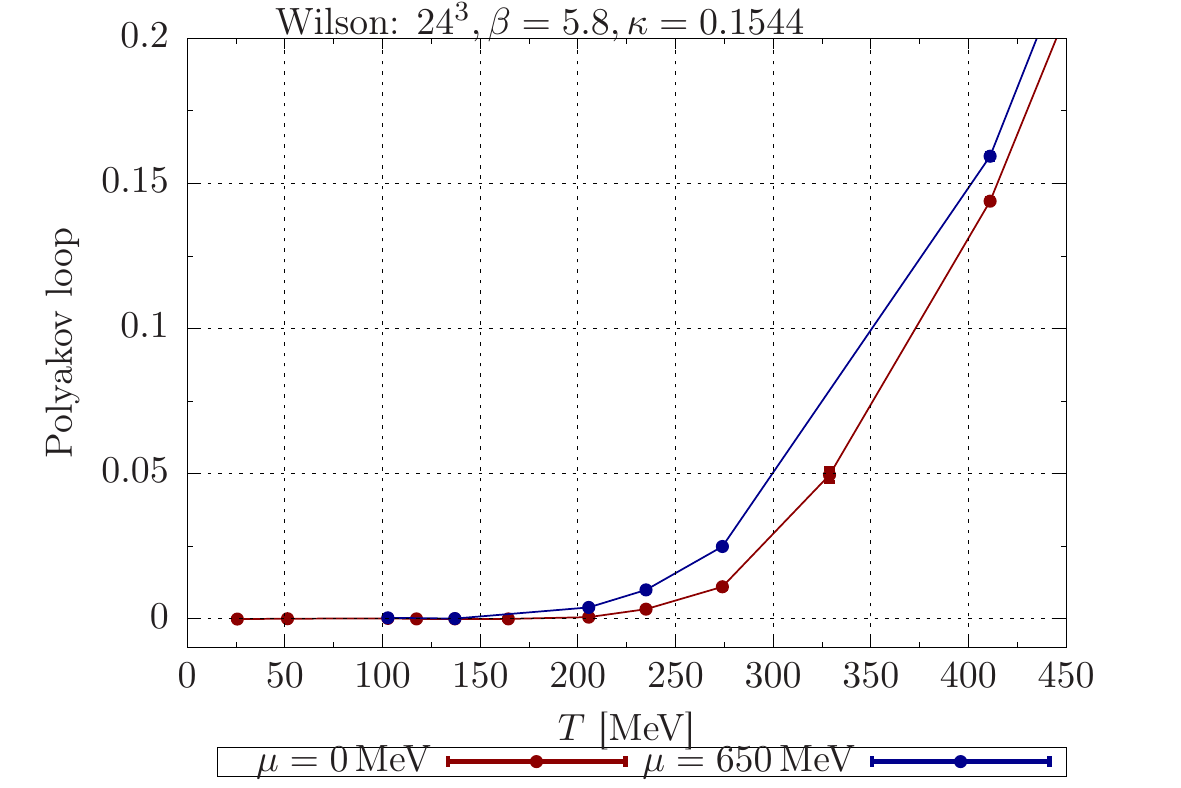}
    \end{minipage}
    \caption{(Left): The Polyakov loop as function of chemical potential $\mu$ for the full range of simulations. As before the gray lines indicate $m_\pi/2$ and $m_N/3$ (Right): The Polyakov loop as function of the temperature $T$ for two choices of $\mu$. }
    \label{fig:poly}
\end{figure}

\begin{figure}[!htb]
    \centering
    \begin{minipage}{.49\textwidth}
    \centering
    \includegraphics[width=0.98\textwidth]{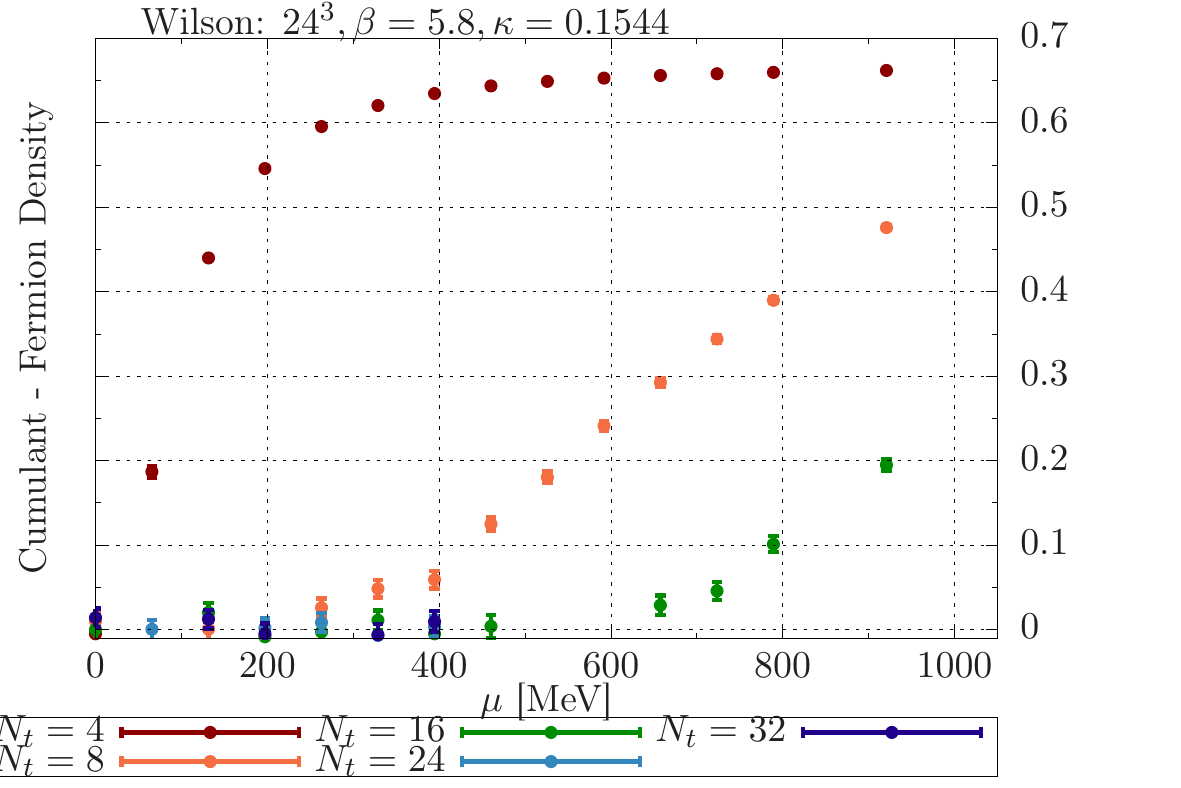}
    \end{minipage}
        \begin{minipage}{.49\textwidth}
    \centering
    \includegraphics[width=0.98\textwidth]{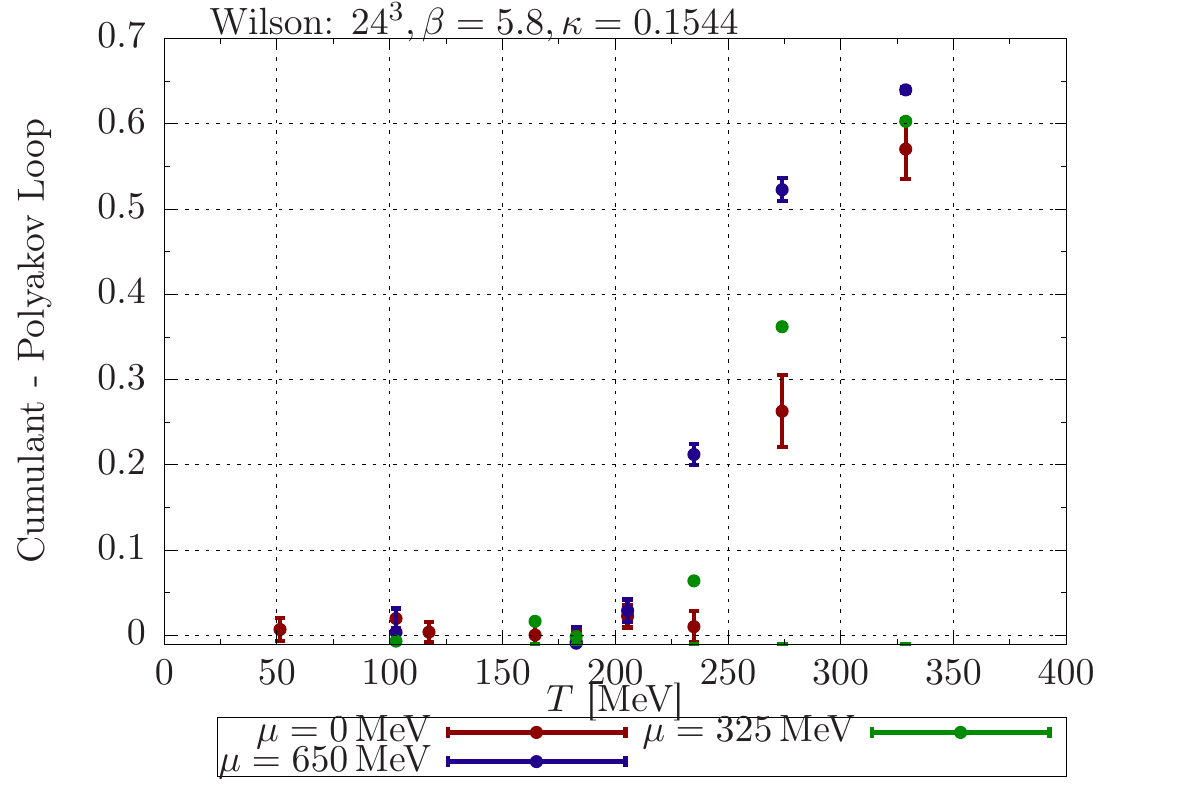}
    \end{minipage}
    
    \caption{The Binder cumulants for the fermion density (left) and the Polyakov loop (right) for different temperatures and chemical potentials. }
    \label{fig:binder}
\end{figure}

To explore the QCD phase diagram, we simulate at different chemical potential and temperatures. Our parameters are chosen such that we span a large region of the QCD phase diagram. To study the phase behaviour, we investigate in particular the fermion density and Polyakov loop. 
\begin{figure}[!htb]
    \centering
    \begin{minipage}{.49\textwidth}
    \centering
    \includegraphics[width=0.98\textwidth]{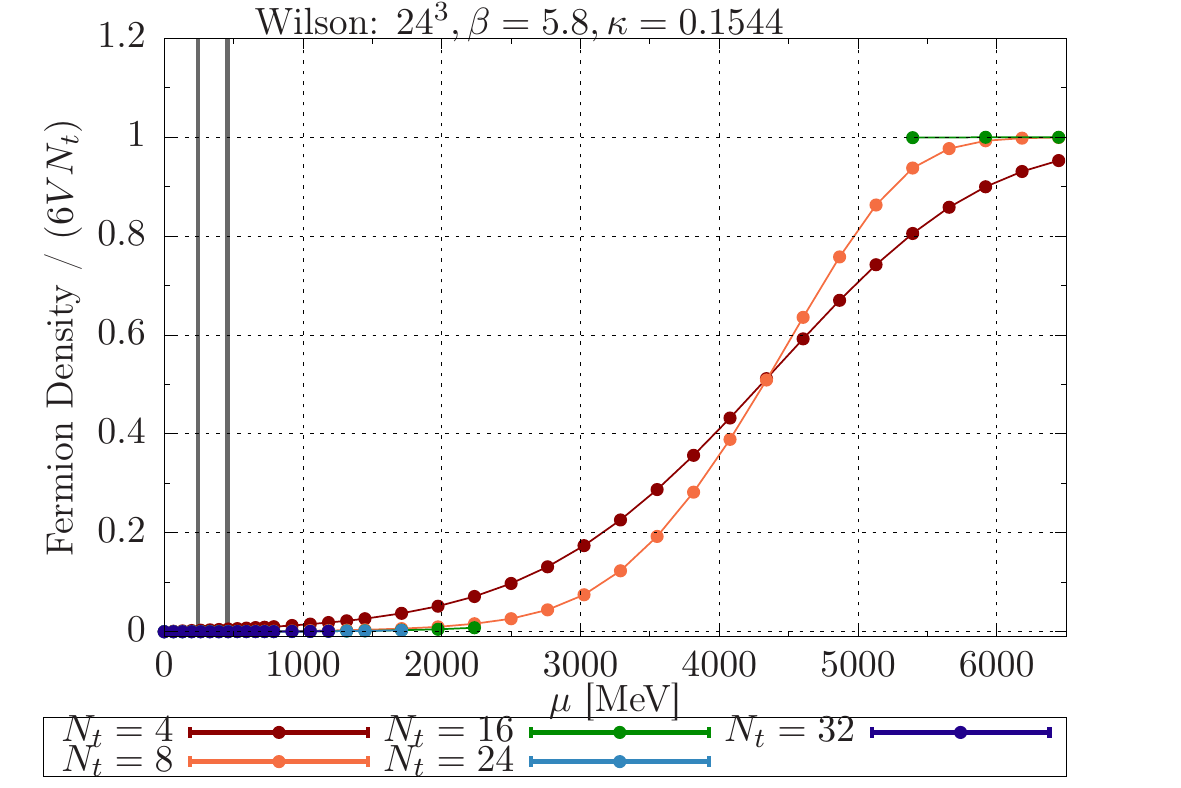}
    \end{minipage}
        \begin{minipage}{.49\textwidth}
    \centering
    \includegraphics[width=0.98\textwidth]{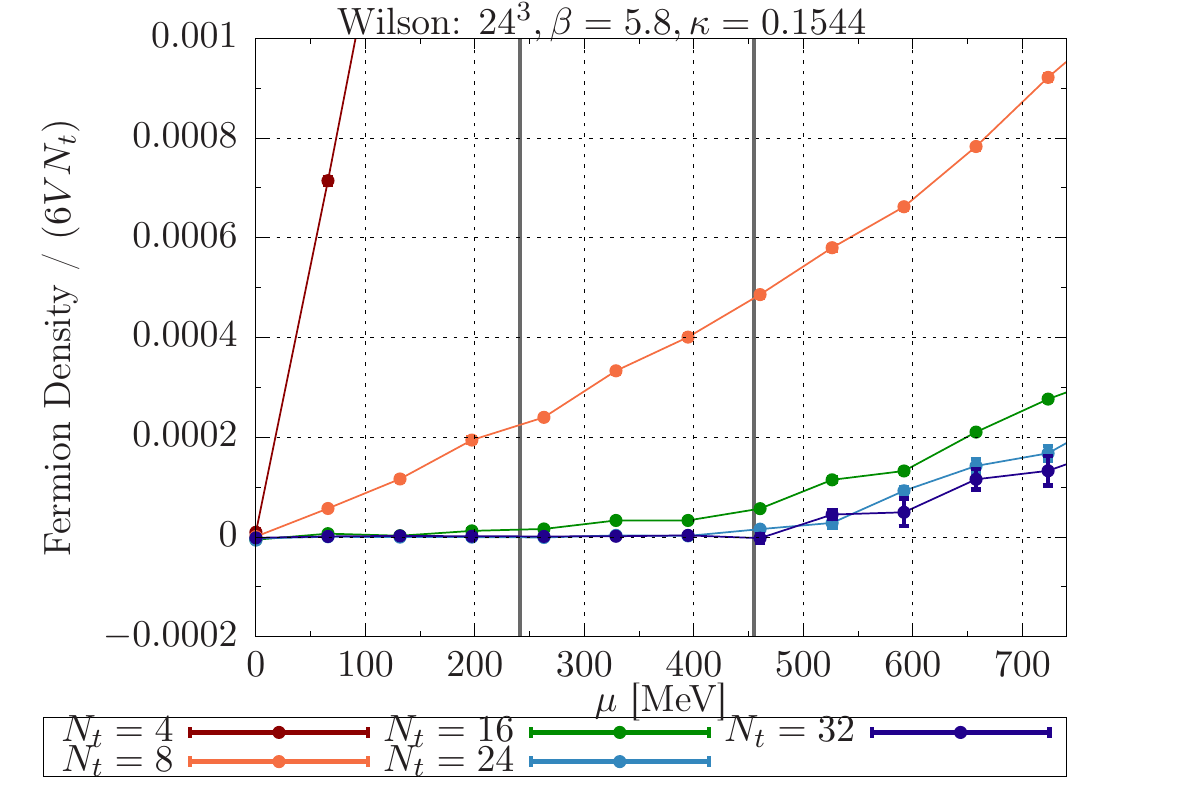}
    \end{minipage}
    \caption{The fermion density normalised over $(6 V N_t)$ for each flavour for large set of lattice extends $N_t$ and chemical potentials $\mu$. The two grey vertical lines correspond to $m_\pi/2$ and $m_N/3$. (Left): The full simulation range. (Right:) Zoom into the view on the small chemical potentials and tiny densities. At low temperatures the fermion density remains zero until $m_N/3$, which is a first sign of the Silver Blaze phenomenon.}
    \label{fig:dens}
\end{figure}
Figure~\ref{fig:dens} shows the fermion density as a function of the chemical potential. Each data point corresponds to an independent complex Langevin simulation. The left panel shows the entire simulation range, whereas the right panel focuses on smaller chemical potentials. For both, the fermion density is divided by $6\, V\, N_t$ for each flavour, i.e. the saturation density. At large $\mu$ we see aforementioned saturation, which is an intrinsic lattice artefact. This occurs when all lattice sites are filled with fermions, so that Pauli blocking does not allow additional fermions to be added to the system. The right-hand side of figure~\ref{fig:dens} also shows an enlarged view of the fermion density for small chemical potentials and tiny values of the density. Interestingly, the fermion density remains $0$ up to the $m_N/3$, which is a manifestation of the so-called Silver Blaze phenomenon~\cite{Cohen:2003kd}. The Polyakov loop is shown in the left panel of figure~\ref{fig:poly}. Here the same lattice artefact is visible for large chemical potentials.
In the right panel of figure~\ref{fig:poly} we show the Polyakov loop as function of the temperature for two chemical potentials. We find that the transition to a nonzero Polyakov loop is occurring at lower temperatures for larger chemical potentials. This allow us to quantify the transition(s). Additionally, we also look at the Binder cumulant~\cite{Binder:1981sa}, defined in the following way
\begin{align*}
    B = 1 - \frac{\langle O^4\rangle}{3 \langle O^2\rangle^2}.
\end{align*}
The Binder cumulant for the fermion density is shown in the left panel of figure~\ref{fig:binder} and for the Polyakov loop in the right panel. The transition to non-zero values for the Binder cumulant occurs at smaller temperatures as the chemical potential increases, as already seen in the right panel of figure~\ref{fig:poly}.

We have covered a large range of temperatures and chemical potentials, however, the simulations at low temperatures become increasingly challenging, from a numerical point of view. In particular, the number of conjugate gradient steps increases sharply as the chemical potential increases, in particular for low temperature simulations.
\begin{figure}[!htb]
    \centering
    \begin{minipage}{.49\textwidth}
    \centering
    \includegraphics[width=0.98\textwidth]{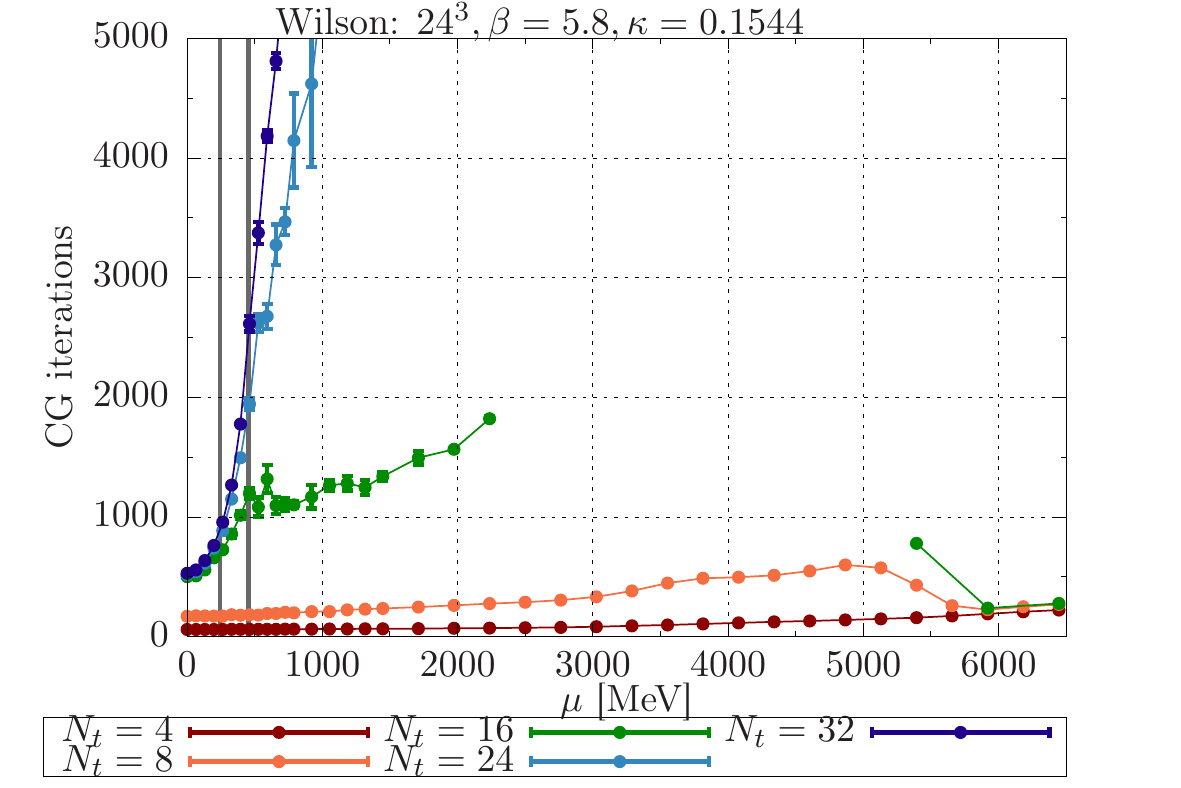}
    \end{minipage}
        \begin{minipage}{.49\textwidth}
    \centering
    \includegraphics[width=0.98\textwidth]{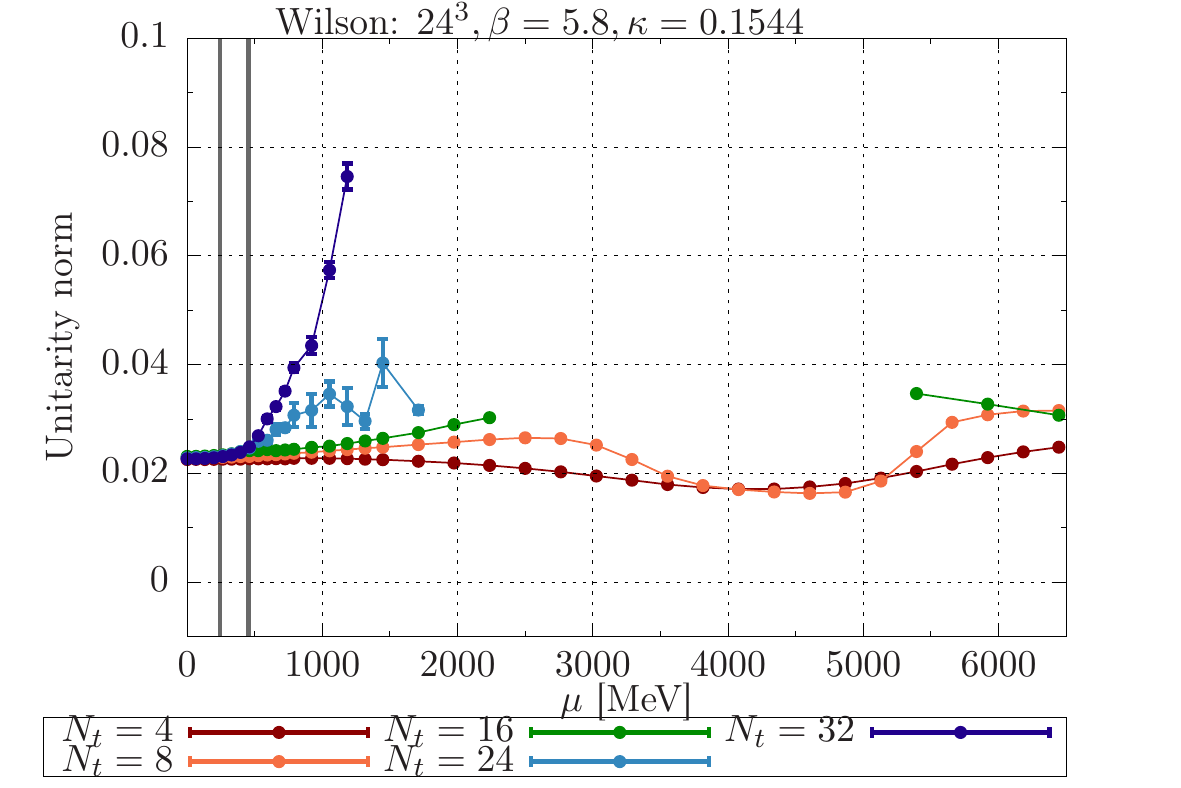}
    \end{minipage}
    
    \caption{(Left): The number of conjugate gradient iterations necessary for the computation of the drift force in the complex Langevin process.  
    (Right): The average unitarity norm of the individual simulation, i.e. the distance to the SU(3) manifold. }
    \label{fig:numerics}
    
\end{figure}
This behaviour can be seen in the left panel of figure~\ref{fig:numerics}. On the right panel, the average unitarity norm is shown, which stays sufficiently small for all simulations. For lower temperatures we can see an increase for larger chemical potentials.

\section{Outlook \& Conclusion}

Here we have reported on our ongoing 
first-principle study of the QCD phase diagram using complex Langevin simulations. We have performed simulations at various chemical potentials and temperatures using two flavour Wilson quarks with a pion mass of approximately $480\,$MeV. In particular, we have found that the fermion density stays zero in the region of $m_\pi/2$ to $m_N / 3$, i.e. the Silver Blaze phenomenon. 

\medskip
In the future, we plan to focus on improved
computations and algorithms, in particular at lower temperatures. The necessary code developments are already underway. Furthermore, we plan to repeat our study with different volumes, in order to quantify the order of the transition(s) by studying the volume scaling.

\acknowledgments
The work of F.A. was supported by the Deutsche Forschungsgemeinschaft (DFG, German Research Foundation) under Germany’s Excellence Strategy EXC2181/1 - 390900948 (the Heidelberg STRUCTURES Excellence Cluster) and under the Collaborative Research Centre SFB 1225 (ISOQUANT). This work was performed using PRACE resources at Hawk (Stuttgart) with project ID 2018194714. This work used the DiRAC Extreme Scaling service at the University of Edinburgh, operated by the Edinburgh Parallel Computing Centre on behalf of the STFC DiRAC HPC Facility (www.dirac.ac.uk). This equipment was funded by BEIS capital funding via STFC capital grant ST/R00238X/1 and STFC DiRAC Operations grant ST/R001006/1. DiRAC is part of the National e-Infrastructure.

\end{document}